\documentclass[twoside,fleqn]{article}
\usepackage{espcrc2}
\usepackage{theorem}

\theoremstyle{break}    \newtheorem{The}{Theorem}
\theoremstyle{plain}    
\theoremstyle{plain}    
\theoremstyle{plain}    
{\theorembodyfont{\rmfamily}     }
{\theorembodyfont{\rmfamily}     }

\def\gsim{{\mathrel{\raise2pt\hbox to 8pt{\raise -5pt\hbox{$\sim$}\hss{$>$}}}}}
\def\rsim{{\mathrel{\raise2pt\hbox to 8pt{\raise -5pt\hbox{$\sim$}\hss{$>$}}}}}
\def\lsim{{\mathrel{\raise2pt\hbox to 8pt{\raise -5pt\hbox{$\sim$}\hss{$<$}}}}}

\begin{document}

\title{
Progress report on the staggered $\epsilon'/\epsilon$ project\thanks{
Presented by W.~Lee.  Research supported in part by BK21, by the SNU
foundation \& Overhead Research fund, by KRF contract
KRF-2002-003-C00033 and by US-DOE contract DE-FG03-96ER40956/A006.}  }
\author{T.~Bhattacharya\address{MS--B285, T-8,
                Los Alamos National Lab, 
		Los Alamos, New Mexico 87545, USA},
        G.T.~Fleming\address{Department of Physics,
                Ohio State University, Columbus, OH 43210, USA},
        G.~Kilcup${}^{\rm b}$,
        R.~Gupta${}^{\rm a}$,
        W.~Lee\address{School of Physics, 
		Seoul National University,
		Seoul, 151-747, South Korea}
        and
        S.~Sharpe\address{Department of Physics,
                University of Washington, 
		Seattle, WA 98195, USA}
}
\begin{abstract}
We report on progress and future plans for calculating kaon weak
matrix elements for $ \epsilon' / \epsilon $ using staggered fermions.
\end{abstract}

\maketitle


The first goal of this project is to provide a check of the results ($
\epsilon' / \epsilon < 0 $) obtained using quenched Domain Wall
fermions (DWF) by the CP-PACS~\cite{ref:CP-PACS:0} and
RBC~\cite{ref:RBC:0} collaborations.
The second goal is to extend the work to dynamical simulations, since
quenching is likely one of the main systematic errors in present
calculations~\cite{ref:wlee:1}.
The third goal is either to find a window for new physics or to
confirm the standard model, when our numerical results are compared
with the observed values.
%

Staggered fermions are appropriate for this purpose, because they
preserve enough chiral symmetry to prevent composite operators from
mixing with wrong chirality operators and to protect quark mass from
additive renormalization, which is essential to calculate $ \epsilon'
/ \epsilon $.
They have the advantages over DWF of requiring less CPU time and
dynamical simulations are already possible with relatively light 
quark masses.
%
%
By construction, staggered fermions carry four degenerate flavors
(also called ``tastes'').
They allow large flavor changing quark-gluon interactions, which makes
it impractical to non-perturbatively determine matching constants for
operators of our interest.
%
%
Another disadvantage is that the unimproved action and operators
receive large perturbative corrections at one loop and have large
scaling violations of order $a^2$.
Both of these disadvantages can be alleviated by improving
staggered fermions using smeared links.
Flavor symmetry breaking in the pion spectrum is significantly reduced
with such links~\cite{ref:orginos:0,ref:hasenfratz:0}.
%
%
%
%

Our project has progressed through the following steps.

\noindent {\bf 1.} We calculated the current-current diagrams for the
gauge-invariant unimproved staggered fermions at one
loop~\cite{ref:wlee:0}, which provided a complete set of perturbative
matching formula for $ \epsilon' / \epsilon $, combined with existing
results for penguin diagrams~\cite{ref:steve:0}.
%

\noindent {\bf 2.} We performed a numerical study on $\epsilon' /
\epsilon$ using the Columbia QCDSP supercomputer~\cite{ref:wlee:1}.
Using the matching formula given in \cite{ref:wlee:0,ref:steve:0}, we
constructed fully one-loop matched gauge-invariant operators.
As expected, we found large perturbative corrections for $B_6$, so
that we cannot quote quantitative results for this although the
statistical uncertainty is under control.
An unexpected result was that different quenched transcriptions of the
continuum operators on the lattice (proposed by Golterman and
Pallante~\cite{ref:golterman:0}) lead to substantially different
values for $B_6$~\cite{ref:wlee:1}.
This indicates a large quenching uncertainty and deserves further
study.
%

\noindent {\bf 3.} The goal was to find an improvement scheme which
can reduce perturbative correction down to 10\% or smaller.
To achieve this goal, we calculated, explicitly, one loop matching
factors for a variety of improved staggered actions and
operators~\cite{ref:wlee:2}: 1) Fat7, 2) Fat7+Lepage, 3) HYP, and 4)
Asqtad-like (Fat7+Lepage+Naik) actions.
We observed that all the above improvement schemes significantly
reduce the size of the matching coefficients.
After a higher level of mean-field improvement, the HYP and Fat7 links
lead to the smallest one-loop corrections~\cite{ref:wlee:2}.
Since the HYP action reduces the non-perturbative flavor symmetry
breaking more efficiently in the pion spectrum, we adopted the HYP
scheme in our numerical study.
%

\noindent {\bf 4.} We studied further on the HYP
link~\cite{ref:wlee:3}.
The HYP links possess some universal properties, which are summarized
in the following 5 theorems.
\begin{The}[SU(3) Projection]
Any fat link can be expanded in powers of gauge fields ($A_\mu$).
\begin{eqnarray*}
& & B_\mu = B_\mu^{(1)} + B_\mu^{(2)} + B_\mu^{(3)} + \cdots
\\ & & B_\mu^{(n)} = {\cal O}(A^n)
\end{eqnarray*}
  \begin{enumerate}
  \item The linear term, $B_\mu^{(1)}$ is invariant under SU(3)
  projection.
  \item The quadratic term, $B_\mu^{(2)}$ is antisymmetric in gauge
  fields.
  \end{enumerate}
\label{theorem:su(3)}
\end{The}
\begin{The}[Triviality of renormalization]
  \begin{enumerate}
  \item At one loop level, only the $B_\mu^{(1)}$ term contributes
    to the renormalization of the gauge-invariant staggered fermion
    operators.
  \item At one loop level, the contribution from $ B_\mu^{(n)}$ for
    any $n \geq 2$ vanishes.
  \item At one loop level, the renormalization of the gauge-invariant
    staggered operators can be done by simply replacing the propagator
    of the $A_\mu$ field by that of the $B_\mu^{(1)}$ field.
  \end{enumerate}
  This theorem is true, regardless of details of the smearing
  transformation.
\label{theorem:renorm}
\end{The}
\begin{The}[Multiple SU(3) projections]
  \begin{enumerate}
  \item The linear gauge field term $B_\mu^{(1)}$ in the perturbative
    expansion is universal.
  \item In general, the quadratic terms may be different from one
    another.  But all of them are antisymmetric in gauge fields.
  \item This theorem is true, regardless of the details of smearing.
  \end{enumerate}
\label{theorem:multi-su(3)}
\end{The}

\begin{The}[Uniqueness]
  If we impose the perturbative improvement condition of removing the
  flavor changing interactions on the HYP action,
  the HYP link satisfies the following:
  \begin{enumerate}
  \item The linear term $B_\mu^{(1)}$ in perturbative expansion is
identical to that of the SU(3) projected Fat7 links.
  \item The quadratic term $B_\mu^{(2)}$ is antisymmetric in gauge
fields.
  \end{enumerate}
\label{theorem:unique}
\end{The}

\begin{The}[Equivalence at one loop]
  If we impose the perturbative improvement condition to remove the
  flavor changing interactions, at one loop level,
  \begin{enumerate}
  \item the renormalization of the gauge invariant staggered operators
    is identical between the HYP staggered action and those improved
    staggered actions made of the SU(3) projected Fat7 links,
  \item the contribution to the one-loop renormalization can be
    obtained by simply replacing the propagator of $A_\mu$ by
    that of $B_\mu^{(1)}$.
  \end{enumerate}
  \label{theorem:equivalence}
\end{The}
The first two theorems were used in~\cite{ref:steve:1}
and~\cite{ref:bernard:0}, although they did not present their
derivations.
For derivations of all five theorems and further details,
see~\cite{ref:wlee:3}.
%
%
%
%
As a result of these theorems, we can prove that for each Feynman
diagram,
\begin{eqnarray}
    \parallel C_{fat} \parallel \ < \ \parallel C_{thin} \parallel \, .
\end{eqnarray}
Here, $ C_{fat} $ ($ C_{thin} $) represents perturbative corrections
to gauge-invariant staggered operators constructed using SU(3)
projected fat links (thin links).
This inequality is not valid for those fat links without SU(3)
projection.
Hence, this lead to a conclusion that we may view the SU(3) projection
of fat links as a tool of tadpole improvement for the staggered
fermion doublers~\cite{ref:wlee:3}.
We also present alternative choices of constructing fat links to
improve staggered fermions in~\cite{ref:wlee:3}.
The above five theorems make the perturbative calculation simpler for
the HYP scheme, because one can perform the calculation merely by
replacing the thin link propagator with that of the HYP links.
This simplicity is extensively used in calculating the renormalization
constants of the four-fermion operators in the next stage.
\begin{table}[t]
\begin{center}
        \begin{tabular}{cc}
        Operators   & $[P \times P][P \times P]_{II}$ \\ \hline
        NAIVE       & $ 2 \times (111.3 - 2 C_N) $    \\
        HYP         & $ 2 \times (8.25 - 2 C_H) $
        \end{tabular}
\end{center}
\caption{One-loop correction to $({\cal O}_3)_{II}$.}
\label{tab:PPPP}
\vspace{-0.3in}
\end{table}
\noindent {\bf 5.} We calculate the current-current diagrams to obtain
perturbative matching coefficients for the staggered four-fermion
operators constructed using the HYP links.
Especially, we are interested in the $({\cal O}_3)_{II}$ operator (we
use the same notation as in~\cite{ref:wlee:0}), because this receives
large perturbative corrections ($\approx 1$) in the case of unimproved
staggered fermions.
\begin{eqnarray*}
({\cal O}_3)_{II} = 
	2 ( [P \times P][P \times P] 
	- [S \times P][S \times P] )_{II}
\end{eqnarray*}
In Tables \ref{tab:PPPP} and \ref{tab:SPSP}, we present values of
one-loop corrections to $({\cal O}_3)_{II}$ constructed using both the
HYP link and unimproved thin link (denoted as NAIVE).
Here, $C_N = 13.159$ and $C_H = 0.7709$, which correspond to tadpole
improvement contributions.
The results shows that, by choosing the HYP scheme, the perturbative
corrections are reduced to $\approx$ 10\% level.
In the case of the HYP scheme, note that the size of one-loop
correction is already under control even without tadpole improvement.
For further details, refer to~\cite{ref:wlee:4}.
\noindent {\bf 6.} At present, we are performing a numerical study
using the Columbia QCDSP supercomputer.
We calculate weak matrix elements for $ \epsilon' / \epsilon $ using
the HYP links at quenched $\beta=6.0$ on the $16^3 \times 64$ lattice.
So far, we have completed measurements on about 80 gauge
configurations.
We plan to complete the one-loop calculations with HYP fermions,
including the penguin diagrams~\cite{ref:wlee:4}.
This will allow us to match the complete set of $ \Delta I = 1/2 $ and
$ 3/2 $ four-fermion operators constructed using HYP links to the
corresponding continuum operators in, say, the NDR scheme.
%

%
%
%
%

%
We have also generalized the definition of the optimal matching scale,
$q^*$~\cite{ref:lepage:0}, to the case of matching factors for
operators with non-vanishing anomalous dimensions~\cite{ref:wlee:4}.
%
%
Ultimately, we wish to extend the calculation to partially quenched
and unquenched QCD.
We thank N.~Christ, C.~Jung, C.~Kim, G.~Liu, R.~Mawhinney and L.~Wu
for their support of this project and assistance with numerical
simulations on the Columbia QCDSP supercomputer.
%

%
%
%
%
%
%
%
\begin{table}[t]
\begin{center}
	\begin{tabular}{cc}
        Operators   & $[S \times P][S \times P]_{II}$ \\ \hline
        NAIVE       & $ 2 \times (95.6 - 6 C_N) $     \\
        HYP         & $ 2 \times (14.6 - 6 C_H) $
        \end{tabular}
\end{center}
\caption{One-loop correction to $({\cal O}_3)_{II}$.}
\label{tab:SPSP}
\vspace{-0.3in}
\end{table}
%
%
%


\end{document}